\documentclass[aps,pre,twocolumn,groupedaddress]{revtex4}

\usepackage{graphicx}
\usepackage{dcolumn}
\usepackage{bm}
\usepackage{epsfig}
\DeclareMathAlphabet{\mathpzc}{OT1}{pzc}{m}{it}

\newcommand{\dummy}
\begin{document}

\title{Kinetics of Phase Separation in Thin Films: Lattice versus 
Continuum Models for Solid Binary Mixtures}

\author{Subir K. Das,$^{1}$ J\"urgen Horbach,$^{2}$ and Kurt Binder$^{3}$}

\affiliation{$^1$Theoretical Sciences Unit, Jawaharlal Nehru Centre for
Advanced Scientific Research, Jakkur, Bangalore 560064, India\\
$^2$Institut f\"ur Materialphysik im Weltraum, Deutsches Zentrum f\"ur
Luft- und Raumfahrt (DLR), 51170 K\"oln, Germany\\
$^3$Institut f\"{u}r Physik, Johannes Gutenberg-Universit\"at
Mainz, Staudinger Weg 7, 55099 Mainz, Germany}
\date{\today}

\begin{abstract}
A description of phase separation kinetics for solid binary (A,B) mixtures
in thin film geometry based on the Kawasaki spin-exchange kinetic Ising
model is presented in a discrete lattice molecular field formulation. It
is shown that the model describes the interplay of wetting layer formation
and lateral phase separation, which leads to a characteristic domain size
$\ell(t)$ in the directions parallel to the confining walls that grows
according to the Lifshitz-Slyozov $t^{1/3}$ law with time $t$ after the
quench. Near the critical point of the model, the description is shown
to be equivalent to the standard treatments based on Ginzburg-Landau
models. Unlike the latter, the present treatment is reliable also at
temperatures far below criticality, where the correlation length in the
bulk is only of the order of a lattice spacing, and steep concentration
variations may occur near the walls, invalidating the gradient square
approximation. A further merit is that the relation to the interaction
parameters in the bulk and at the walls is always transparent, and the
correct free energy at low temperatures is consistent with the time
evolution by construction.
\end{abstract}

\maketitle

\section{Introduction}
A basic problem of both materials science \cite{1,2} and statistical
mechanics of systems out of equilibrium \cite{2,3,4,5} is the process
of spinodal decomposition of binary (A,B) mixtures.  When one brings
the system by a sudden change of external control parameters (e.g., a
temperature quench) from an equilibrium state in the one-phase region of
the mixture to a state inside of the miscibility gap, thermal equilibrium
requires coexistence of macroscopically large regions of A-rich and B-rich
phases. Related phenomena also occur in systems undergoing order-disorder
phase transitions, and it is also possible that an interplay of ordering
and phase separation occurs, particularly in metallic alloys with complex
phase diagrams \cite{1,2,6}.

The recent interest in nanostructured materials and thin films has led to
consider the effects of walls or free surfaces on the kinetics of such
phase changes \cite{7,8,9,10,11,12,13,14,15,16,17,18,19}. In a binary
mixture, it is natural to expect that one of the components (say, A)
will be preferentially attracted to the surface. Already in the one phase
region, this attraction will lead to the formation of surface enrichment
layers of the preferred component at the walls, but the thickness of these
layers will be small, viz., of the order of the correlation length $\xi$
of concentration fluctuations in the mixture \cite{20}. Only near the
critical point this surface enrichment becomes long ranged \cite{21,22}.
For temperatures below the critical point, however, the behavior is
more complicated due to the interplay between bulk phase separation
and wetting phenomena \cite{23,24,25,26,27,28,29}. In a semi-infinite
geometry, a B-rich domain of a phase separated mixture would be coated
with an A-rich wetting layer at the surface, if the temperature is
above the wetting transition.  Of course, for thin films the finite
film thickness $D$ also constrains the growth of wetting layers: E.g.,
for short range forces between the walls and the A-atoms the equilibrium
thickness of a ``wetting layer'' is of the order of $\xi \ln (D/\xi)$
\cite{29,30,31}, while at lower temperatures where the surface is nonwet
the thickness of the surface enrichment layer again only is of the order
of $\xi$. Also the phase diagram of the system in a thin film geometry
differs from that of the bulk, in analogy to capillary condensation of
fluids \cite{29,31,32,33,34}, and very rich phase diagrams may occur,
in particular, if the surfaces confining the thin film are not of the
same type \cite{29,35,36,37}. These changes in the phase behavior of
thin films are reflected in the kinetics of phase separation in these
systems \cite{18,19}, of course.

Despite the large theoretical activity on surface effects on phase
separation kinetics \cite{8,9,10,12,13,14,15,17,18,19} the applicability
of these works on experiments is very restricted.  Actually, most of the
experiments deal with thin films of fluid binary mixtures \cite{7,11,16},
but almost all theoretical works \cite{8,9,10,12,13,14,15,17,18} deal with
``model $B$'' \cite{38}, where hydrodynamic interactions are neglected,
and hence this model is really appropriate only for solid mixtures.

A second restriction on the applicability of the theory is the
fact that it is based on the time-dependent Ginzburg-Landau model
\cite{1,2,3,4,5,6}, and thus is applicable only when the correlation
length $\xi$ is very large, i.e.~in the immediate vicinity of the bulk
critical point. For a thin film with short range surface forces, the
``standard model'' is based on a free energy functional of an order
parameter $\psi$, and this functional consists of a bulk term ($F_{\rm
b}$) and two terms representing the two surfaces $\mathcal{S}1$,
$\mathcal{S}2$ \cite{8,10,12,13,14,17,18}, $F[\psi]=F_{\rm b} +
F_{\mathcal{S}1} + F_{\mathcal{S}2}$, with
\begin{eqnarray}
F_{\rm b} [\psi] & = & 
\int d \vec{r} \; \Big\{-\frac{\psi(\vec r)^2}{2} +
\frac{\psi(\vec r)^4}{4} + \frac{1}{4} 
[\vec{\nabla} \psi(\vec r)]^2 \Big\},
\label{eq1} \\
F_{\mathcal{S}1} & = & 
\int\limits_{\mathcal{S}1} d \vec{\rho} \; \Big\{-\frac{g}{2} 
[\psi (\vec{\rho}, 0)]^2 -h_{\mathcal{S}1} \psi(\vec{\rho}, 0) 
\nonumber \\
& & - \gamma \psi(\vec{\rho}, 0) 
\frac{\partial \psi}{\partial z} |_{z=0} \Big \},
\label{eq2} \\
F_{\mathcal{S}2} & = & 
\int\limits_{\mathcal{S}2} d \vec{\rho} \; \Big\{-\frac{g}{2}
[\psi (\rho, D)\big]^2
-h_{\mathcal{S}2} \psi(\vec{\rho}, D) 
\nonumber \\
& & + \gamma \psi(\vec{\rho}, D)
\frac{\partial \psi}{\partial z} |_{z=D} \Big \}. \label{eq3}
\end{eqnarray}
Here $\psi(\vec{r})$ is the order parameter which is proportional to the
density difference between the two species; it is normalized such that the
coexisting A-rich and B-rich bulk phases for temperature $T < T_{\rm cb}$
($T_{\rm cb}$ being the bulk critical temperature) correspond to $\psi
= \pm 1$, respectively. All lengths are measured in units of $2\xi$,
with $\xi$ denoting the bulk correlation length at the coexistence
curve. The terms $F_{\mathcal{S}1}$ and $F_{\mathcal{S}2}$ are the local
contributions from the surfaces $\mathcal{S}1$ and $\mathcal{S}2$, which
for a film of thickness $D$, are located at $z=0$ and $z=D$, respectively,
orienting the $z$-axis perpendicular to the surfaces, while $\vec{\rho}$
denotes the $(d-1)$ coordinates parallel to the surfaces, $d$ being the
dimensionality. In $F_{\mathcal{S}1}$ there are parameters $g, \gamma$,
and $h_{\mathcal{S}1}$, which can be related to the temperature $T$
and various parameters of an Ising ferromagnet with a free surface at
which a surface magnetic field $H_{\mathcal{S}1}$ acts \cite{8,39}
\begin{eqnarray} \label{eq4}
h_{\mathcal{S}1} & = & 
4(H_{\mathcal{S}1}/T)(\sqrt{12})^3 \xi^5/\sqrt{3},\nonumber\\
g & = & 8(4J_s/J-5)\xi^4,\nonumber\\
\gamma & = & 4 \xi ^3,
\end{eqnarray}
with
\begin{equation} \label{eq5}
\xi=\frac{1}{\sqrt{12}} \Big[1-\frac{T}{T_{\rm cb}} \Big]^{-1/2}, \,
\, k_{\rm B}T_{\rm cb}= 6J. \quad 
\end{equation}
In Eqs.~(\ref{eq4},~\ref{eq5}), a simple cubic lattice was assumed, the
$z$-axis coinciding with a lattice axis. Nearest neighbor Ising spins in
the lattice interact with an exchange coupling $J$, except for the surface
plane at $z=0$ where the exchange coupling is $J_{\mathcal{S}}$. For
deriving Eqs.~(\ref{eq1},~\ref{eq2}) from a layerwise molecular field
approximation, the limit $\xi \rightarrow \infty$ needs to be taken,
with $\psi (\vec{\rho}, z)=m_n (\vec{\rho}, t)/m_{\rm b}$, where $n$
is the layer index of the lattice, $n=1$ being the surface plane, and
$m_{\rm b}$ [$=\sqrt{3} (1-T/T_{\rm cb})^{1/2}$] the bulk magnetization
(note that $m_{\rm b} \rightarrow 0$ in the considered limit). Finally,
$F_{\mathcal{S}2}$ describes analogously, the surface free energy of
the surface at $z=D$, with $h_{\mathcal{S}2}$ being related to the
surface field $H_{\mathcal{S}2}$ analogously to Eq.~(\ref{eq4}). In
Eq.~(\ref{eq5}), $\xi$ is measured in units of the lattice spacing $a$
of the molecular field lattice model, which often, for the sake of
convenience, we set to 1.

Dynamics is associated to the model assuming that in the bulk the order
parameter evolves according to the (nonlinear) Cahn-Hilliard equation
\cite{1,2,3,4,5,6}
\begin{equation} \label{eq6}
\frac{\partial}{\partial \tau} \psi (\vec{r}, \tau) =-\vec{\nabla} \cdot
\vec{J} (\vec{r}, \tau)= \vec{\nabla} \cdot \Big[\vec{\nabla}
\Big(\frac{\delta F}{\delta \psi} \Big) \Big], \quad
\end{equation}
while the surfaces amount to two boundary conditions each \cite{20},
which can be written as
\begin{eqnarray}
\tau_0 \frac{\partial}{\partial \tau} \psi(\vec{\rho}, 0,\tau) 
& = & - \frac{\delta \mathcal{F}}{\delta \psi(\vec{\rho}, 0,\tau)} 
\nonumber\\
& = & h_{\mathcal{S}1} + g \psi(\vec{\rho}, 0, \tau) 
+ \gamma \frac{\partial \psi}{\partial z} |_{z=0}, \label{eq7} \\
J_z(\vec{\rho}, 0, \tau) & = &
- \frac{\partial}{\partial z}
\left[-\psi + \psi^3 -\frac{1}{2} \nabla^2 \psi \right]  
= 0. \label{eq8}
\end{eqnarray}
Equation~(\ref{eq7}) describes a non-conserved relaxation (``model A''
\cite{38}) for the order parameter at the surface, $\tau_0$ setting the
time scale. Since $\psi(\vec{\rho}, 0, \tau)$ relaxes much faster than
the time scales of phase separation away from the surface, one may put
$\tau_0=0$. The equations for $z=D$ are fully analogous to those for
$z=0$. We emphasize that Eq.~(\ref{eq6}) can also be derived \cite{40}
from a continuum approximation to a molecular field approximation
to a description of a Kawasaki kinetic Ising model \cite{41},  and
Eqs.~(\ref{eq7},~\ref{eq8}) can be derived from a Kawasaki model with
a free surface as well \cite{20}. However, it is clear that the model,
Eqs.~(\ref{eq6},~\ref{eq7},~\ref{eq8}), can only represent the molecular
field Kawasaki kinetic Ising model accurately when $\xi \gg$ one lattice
unit. This fact was already discussed in \cite{20} in the one-phase
region, where the linearized molecular field equations on the lattice were
solved to describe the kinetics of surface enrichment for $T > T_{\rm
cb}$, and it was found that the lattice and continuum theories agree
when approximations such as $\exp (-1/\xi) \simeq 1-1/\xi$ become valid.

Thus, the model Eqs.~(\ref{eq6},~\ref{eq7},~\ref{eq8}) can describe a
solid binary mixture accurately near the bulk critical point only: far
below $T_{\rm cb}$, terms of order $\psi^6$ and higher would be needed in
Eq.~(\ref{eq1}) already, and when $\xi$ is of the order of the lattice
spacing $a$, also higher order gradient terms $(\vec{\nabla}^2 \psi)^2$
etc. would be required for an accurate continuum description. However,
the numerical solutions of Eqs.~(\ref{eq6},~\ref{eq7},~\ref{eq8})
require anyway a discretization: Often a mesh size $\Delta x = \Delta y=
\Delta z=1$ is used to solve these equations \cite{8,10,12,13,14,17,18}.
This essentially corresponds to a lattice with spacing $2\xi$, rather
than the spacing $a$ of the underlying Ising lattice. Having in mind
applications of the theory at temperatures that are not close to the
critical point, however, $\xi$ is of the order of $1$.

Thus, it is plausible that one could solve with a comparable effort the
original (nonlinear) molecular field equations for the Kawasaki kinetic
Ising model on the lattice, that are underlying this theory. The advantage
of such a treatment clearly would be that the model has a well-defined
microscopic meaning at all temperatures. Near the critical temperature,
the results of this treatment should become indistinguishable from the
solution of Eqs.~(\ref{eq6},~\ref{eq7},~\ref{eq8}), of course.

The purpose of the present work is to show that indeed such a lattice
mean field approach to spinodal decomposition in thin films (and in
the bulk) is both feasible and efficient. In Sec.~II, we present the
discrete lattice analogue of Eqs.~(\ref{eq1},~\ref{eq2},~\ref{eq3}),
and (\ref{eq6},~\ref{eq7},~\ref{eq8}), following up on the work
by Binder and Frisch \cite{20}. In Sec.~III, we present various
numerical results for deep quenches (i.e., temperatures far below
criticality) and discuss the resulting structure formation for a
symmetric film, both in directions parallel and perpendicular to the
walls. In Sec.~IV, we present a comparison to the continuum approach,
Eqs.~(\ref{eq6},~\ref{eq7},~\ref{eq8}).  Finally, Sec.~V summarizes
our paper, containing also an outlook on simulations where the basic
atomistic model is not a lattice model, as is the case for spinodal
decomposition in fluid binary mixtures.

\section{Molecular Field Theory for the Kawasaki Kinetic Ising
Model in a Thin Film}

The system that is considered is the ferromagnetic Ising model with
nearest neighbor exchange on the simple cubic lattice, described by
the Hamiltonian
\begin{eqnarray} \label{eq9}
\mathcal{H}& = & -J \sum\limits_{\langle i,j \rangle \atop {\rm
bulk}} S_i S_j - J_s \sum \limits_{\langle i,j \rangle \,
\atop  \mathcal{S}1, \mathcal{S}2} S_i S_j \nonumber \\
& & - H \sum \limits_i S_i - H_{\mathcal{S}1} \sum\limits_{i \, \in
\, \mathcal{S}1 } S_i  - H_{\mathcal{S}2} \sum\limits_{i
\,\, \in \, \mathcal{S}2} S_i. \,\,
\end{eqnarray}
Here $S_i=\pm 1$, lattice sites are labeled by the index $i$, and
the first sum runs over all nearest neighbor pairs except those in
the surfaces $\mathcal{S}1$ and $\mathcal{S}2$.  Note that now for a
film of thickness $D$, the surfaces $\mathcal{S}1$ and $\mathcal{S}2$
are located at $n=1$ and $n=n_{\mbox{max}}=D+1$, $n$ being an index
labeling the lattice planes in the $z$-direction perpendicular to the
surfaces. Also note that all distances will be measured in units of
the lattice constant $a$.  The term $-H\sum\limits_i S_i$ describes
the Zeeman energy, $H$ being the bulk magnetic field, and the sum runs
over all sites of the lattice. It should be remembered that in the
interpretation of the Ising model related to binary mixtures, spin up
corresponding to $A$ and spin down to $B$, $H$ would correspond to a
chemical potential difference between the species.

The molecular field equations for the local magnetization $m_n
(\vec{\rho})=\langle S_i \rangle$ (we denote the index $i$ of a
lattice site by the index $n$ of the plane to which it belongs and
a coordinate $\vec{\rho}$ in this plane) become \cite{20}
\begin{eqnarray}
m_n(\vec{\rho}) & = & \tanh \frac{J}{k_{\rm B}T} \Big[m_{n+1} (\vec{\rho})
+ m_{n-1} (\vec{\rho}) \nonumber\\
& & + \sum_{\Delta \vec{\rho}}
m_n (\vec{\rho} + \Delta \vec{\rho}) + \frac{H}{J} \Big], \quad
2 \leq n \leq D, \label{eq10} \\
m_1(\vec{\rho}) & = & 
\tanh \frac{J}{k_{\rm B}T} \Big[m_2 (\vec{\rho}) + \frac{J_S}{J} 
\sum\limits_{\Delta \vec{\rho}} m_1(\vec{\rho}+\Delta\vec{\rho}) \nonumber \\ 
& & + \frac{H_{\mathcal{S}1}+H}{J} \Big], \quad n=1, \label{eq11} \\
m_{D+1}(\vec{\rho}) & = & 
\tanh \frac{J}{k_{\rm B}T} \Big[m_{D} (\vec{\rho})
+ \frac{J_S}{J} \sum\limits_{\Delta \vec{\rho}} m_{D+1} (\vec{\rho} +
\Delta \vec{\rho}) \nonumber\\
& & + \frac{H_{\mathcal{S}2} + H}{J} \Big], \quad n=D+1. \,
\label{eq12}
\end{eqnarray}
Here $\Delta \vec{\rho}$ is a vector connecting site $i$ with one of
its 4 nearest neighbors in a layer.  In the bulk, Eq.~(\ref{eq10})
reduces to the well-known transcendental equation
\begin{equation} \label{eq13}
m_{\rm b} = \tanh [\frac{J}{k_{\rm B}T} (6m_{\rm b} + H/J) ]  ,
\end{equation}
from which $T_{\rm cb}= 6J/k_{\rm B}$ straightforwardly follows. 

We also note that Eqs.~(\ref{eq10},\ref{eq11},\ref{eq12}) correspond to
the free energy
\begin{equation} \label{eq14}
F = \sum\limits^{D+1}_{n=1} (E_n - TS_n)
\end{equation}
with
\begin{eqnarray} \label{eq15}
\frac{S_n}{k_{\rm B}T} & = &
\sum\limits_{\vec{\rho}} \Big \{\Big(\frac{1 + m_n(\vec{\rho})}{2}\Big) 
\ln \Big( \frac{1 + m_n(\vec{\rho})}{2}\Big) \nonumber\\
& & + \frac{1-m_n(\vec{\rho})}{2} \ln \Big(\frac{1 -m_n(\vec{\rho})}{2} 
\Big) \Big\}, \label{eq15} \\
E_n & = & 
- \sum\limits_{\vec{\rho}}\Big \{m_n(\vec{\rho}) H +
\frac{J}{2} m_n(\vec{\rho}) (m_{n-1} (\vec{\rho}) + m_{n+1}(\vec{\rho})) 
\nonumber\\
& & + \frac{1}{2} J m_n(\vec{\rho}) \sum\limits_{\Delta \vec{\rho}}
m_n(\vec{\rho} + \Delta \vec{\rho})\Big \}, 
\quad  2 \leq n \leq D, \label{eq16} \\
E_1 & = & 
- \sum\limits_{\vec{\rho}} \{m_1(\vec{\rho}) (H + H_1) +
\frac{1}{2} J m_1(\vec{\rho}) m_2(\vec{\rho}) \nonumber \\
& & + \frac{1}{2} J_S m_1(\vec{\rho}) \sum\limits_{\Delta
\vec{\rho}} m_1(\vec{\rho} + \Delta \vec{\rho}) \Big \}, 
\quad n=1, \label{eq17} \\
E_{D+1} & = & 
- \sum\limits_{\vec{\rho}} \Big\{ m_{D+1}(\vec{\rho}) (H + H_{\mathcal{S}2})
+ \frac{J}{2} m_{D+1}(\vec{\rho}) m_{D}(\vec{\rho}) \nonumber\\
& & + \frac{J_{\mathcal{S}}}{2} m_{D+1}(\vec{\rho}) 
\sum\limits_{\Delta \vec{\rho}} (\vec{\rho} + \Delta \vec{\rho}) \Big\}, 
\quad n=D+1. \label{eq18}
\end{eqnarray}
Of course, Eqs.~(\ref{eq10},~\ref{eq11},~\ref{eq12}) can be derived from
Eq.~(\ref{eq14}) via
\begin{equation} \label{eq19}
\Big( \frac{\partial F}{\partial m_n(\vec{\rho})}
\Big)_{T,H,H_1, \{m_n(\vec{\rho}) \}'} = 0
\end{equation}
The prime on $\{m_n(\vec{\rho})\}'$ indicates that all local
magnetizations are held constant except the one that appears in the
considered derivative. Of course, in equilibrium there is no explicit
dependence on $\vec{\rho}$, although there clearly is a dependence on $n$.

However, already in equilibrium the $\vec{\rho}$-dependence is
useful, when it is understood that the bulk field has a $\vec
\rho$-dependence. Considering a wave vector dependent bulk field,
one then can derive from Eq.~(\ref{eq10}) for $D \rightarrow \infty$
the wave vector dependent susceptibility $\chi(\vec{k})$. By linear
response to the wave vector dependent field one finds, as is well-known,
\begin{equation} \label{eq20}
\chi(\vec{k})= \frac{1}{k_{\rm B}T} \; \frac{1 -
m_{\rm b}^2}{1-[J(\vec{k})/k_{\rm B}T ] (1-m_{\rm b}^2)}, \quad 
\end{equation}
where $J(\vec{k})$ is the Fourier transform of the exchange interaction,
i.e., in our case
\begin{equation} \label{eq21}
J(\vec{k}) = 2J [\cos (k_xa) + \cos (k_ya) + \cos (k_za)] \simeq 6 J
- J k^2a^2. \, 
\end{equation}
From Eqs.~(\ref{eq20},~\ref{eq21}) we readily see that
\begin{equation} \label{eq22}
\chi(\vec{k}) =\chi (0) / [ 1 + k^2 \xi^2a^2] , \quad \chi (0) =1 /
\Big[\frac{k_{\rm B}T}{1-m_{\rm b}^2} - k_{\rm B} T_{\rm cb} \Big]
\end{equation}
and
\begin{equation} \label{eq23}
\xi^2 = \frac{J}{\frac{k_{\rm B}T}{1 - m^2_b} - k_{\rm B}T_{\rm cb}} .
\end{equation}
Equation (\ref{eq23}) reduces to the result quoted in Eq.~(\ref{eq5})
when $ T \rightarrow T_{\rm cb}$, but also shows that $\xi \rightarrow 0$
when $T \rightarrow 0$.

As discussed in \cite{20}, we associate dynamics to the Ising model via
the Kawasaki spin exchange model \cite{41}, considering exchanges between
nearest neighbors only. Following the method of \cite{40}, one obtains
with the help of the Glauber \cite{42} transition probability a set of
coupled kinetic equations for the local time-dependent mean magnetizations
$\langle S_i (t) \rangle \equiv m_n (\vec{\rho}, t)$ from the (exact)
master equation \cite{41} in molecular field approximation. Denoting
the time scale in the transition probability as $\tau_S$, this set of
equations is:
\begin{itemize}
\item[(i)] $3 \leq n \leq D-1$ (bulk case)
\begin{eqnarray} \label{eq24}
&& 2\tau_S \frac{d}{dt} m_n (\vec{\rho}, t)=-6 m_n (\vec{\rho},
t)\nonumber\\
&& + m_{n-1} (\vec{\rho}, t) + m_{n +1} (\vec{\rho}, t)\nonumber\\
&& + \sum\limits_{\Delta \vec{\rho}} m_n(\vec{\rho} + \Delta
\vec{\rho}, t)\nonumber\\
&& + [1-m_n (\vec{\rho} , t) m_{n-1} (\vec{\rho}, t)] \tanh
\frac{J}{k_{\rm B}T} [m_{n +1} (\vec{\rho}, t)\nonumber\\
&& + m_{n-1}(\vec{\rho}, t) + \sum\limits_{\Delta \vec{\rho}} m_n
(\vec{\rho} + \Delta \vec{\rho}, t)\nonumber\\
&& - m_n (\vec{\rho},t) -m_{n-2}(\vec{\rho}, t) -
\sum\limits_{\Delta \vec{\rho}} m_{n-1}
(\vec{\rho} + \Delta \vec{\rho}, t)] \nonumber\\
&&+ [1-m_n (\vec{\rho}, t) m_{n+1} (\vec{\rho}, t)] \tanh
\frac{J}{k_{\rm B}T} [m_{n+1}(\vec{\rho}, t) \nonumber\\
&& + m_{n-1} (\vec{\rho},t) + \sum\limits_{\Delta
\vec{\rho}} m_n (\vec{\rho} + \Delta \vec{\rho}, t)-m_n(\vec{\rho}, t) \nonumber\\
&& -m_{n+2} (\vec{\rho},t)-\sum\limits_{\Delta \vec{\rho}}
m_{n+1} (\vec{\rho} + \Delta \vec{\rho},t)] \nonumber\\
&&+ \sum\limits_{\Delta \vec{\rho}} [1-m_n (\vec{\rho}, t)
m_n (\vec{\rho} + \Delta \vec{\rho}, t)]\nonumber\\
&& \tanh \frac{J}{k_{\rm B}T} [m_{n+1} (\vec{\rho}, t) + m_{n-1}
(\vec{\rho}, t) +
\nonumber\\
&& \sum\limits_{\Delta \vec{\rho}\;'} m_n (\vec{\rho} + \Delta
\vec{\rho}\;', t) - m_{n+1} (\vec{\rho} + \Delta \vec{\rho},
t)\nonumber\\
&& - m _{n-1} (\vec{\rho} + \Delta \vec{\rho}, t)\nonumber\\
&&-\sum\limits_{\Delta \vec{\rho}\;'} m_n (\vec{\rho} + \Delta
\vec{\rho} + \Delta \vec{\rho}\;', t)].
\end{eqnarray}
Factors such as $[1-m_n m_{n-1}]$ arise from the mean field
approximations to factors $(1-S_i S_j)$ that express the fact that
an exchange of $S_i$ with $S_j$ changes the magnetization $S_i$
only if $S_i$ and $S_j$ are oppositely oriented. In
Eq.~(\ref{eq24}), exchanges of a spin at site $\vec{\rho}$ in
layer $n$ with spins in layers $n-1$, $n + 1$, and the same layer
$n$ need to be considered. In the argument of the $\tanh$
functions, the difference of the effective fields acting on the
spins that are exchanged is found. One can verify that
Eq.~(\ref{eq24}) reduces to Eq.~(\ref{eq10}) if $d m_n
(\vec{\rho}, t)/ dt =0$ is assumed: in equilibrium, the exchange
of a spin with another one is exactly compensated by the inverse
process.
The kinetic equations near the wall are similar; one has to
consider that in layer 1 a field $H_{\mathcal{S}1}$ is acting, and that no spin
exchange into the layer $n=0$ (the wall) is possible. Hence, for

\item[(ii)] $n=2$
\begin{eqnarray} \label{eq25}
&& ~~~2\tau_S \frac{d}{dt} m_2 (\vec{\rho}, t) = - 6 m_2 (\vec{\rho},
t) + m_1 (\vec{\rho}, t)\nonumber\\
&& + m_3 (\vec{\rho}, t)+
\sum\limits_{\Delta \vec{\rho}} m_2 (\vec{\rho} + \Delta \vec{\rho}, t) \nonumber\\
&& + [1-m_2 (\vec{\rho}, t) m_1 (\vec{\rho}, t)] \tanh
\frac{J}{k_{\rm B}T} [m_3 (\vec{\rho}, t) \nonumber\\
&&  + m_1 (\vec{\rho}, t) + \sum\limits_{\Delta \vec{\rho}} m_2
(\vec{\rho} + \Delta
\vec{\rho}, t)\nonumber\\
&&- \frac{H_{\mathcal{S}1}}{J} -m_2 (\vec{\rho}, t) - \frac{J_S} {J}
\sum\limits_{\Delta \vec{\rho}} m_1 (\vec{\rho} + \Delta
\vec{\rho},t)] \nonumber\\
&&+ [1-m_2 (\vec{\rho}, t) m_3 (\vec{\rho}, t)] \tanh
\frac{J}{k_{\rm B}T} [m_3 (\vec{\rho}, t) + m_1 (\vec{\rho}, t)
\nonumber\\
&& + \sum\limits_{\Delta \vec{\rho}} m_2 (\vec{\rho} + \Delta
\vec{\rho}, t) \nonumber\\
&& - m_4 (\vec{\rho}, t) - m_2 (\vec{\rho}, t) -
\sum\limits_{\Delta \vec{\rho}} m_3 (\vec{\rho} + \Delta
\vec{\rho}, t)] \nonumber\\
&& + \sum\limits_{\Delta \vec{\rho}} [1-m_2 (\vec{\rho}, t) m_2
(\vec{\rho} + \Delta \vec{\rho}, t)] \nonumber\\
&& \tanh \frac{J}{k_{\rm B}T} [m_3 (\vec{\rho}, t) + m_1 (\vec{\rho}, t)
+\sum\limits_{\Delta \vec{\rho}} m_2 (\vec{\rho} + \Delta \vec{\rho}, t)]\nonumber\\
&& - m_3 (\vec{\rho} + \Delta \vec{\rho}, t) - m_1 (\vec{\rho} +
\Delta \vec{\rho}, t)\nonumber\\
&& - \sum\limits_{\Delta \vec{\rho}\;'} m_2 (\vec{\rho} + \Delta
\vec{\rho} + \Delta \vec{\rho}\;',t),
\end{eqnarray}
and for

\item[(iii)] $n=1$ (now only 5 neighbors are available for an
exchange)\\
\begin{eqnarray} \label{eq26}
&& 2 \tau_S \frac{d} {dt} m_1 (\vec{\rho}, t) = - 5m_1
(\vec{\rho}, t) + m_2 (\vec{\rho}, t)\nonumber\\
&& + \sum\limits_{\Delta \vec{\rho}} m_1 (\vec{\rho} + \Delta
\vec{\rho}, t) + \nonumber\\
&& [1-m_1 (\vec{\rho}, t) m_2 (\vec{\rho}, t)] \tanh
\frac{J}{k_{\rm B}T} [m_2 (\vec{\rho}, t) \nonumber\\
&&  + \frac{H_{\mathcal{S}1}}{J} + \frac{J_{\mathcal{S}}}{J} \sum\limits_{\Delta \rho} m_1
(\vec{\rho} + \Delta \vec{\rho},
t)\nonumber\\
&& - m_3 (\vec{\rho}, t) - m_1 (\vec{\rho}, t) -
\sum\limits_{\Delta \vec{\rho}} m_2 (\vec{\rho} + \Delta
\vec{\rho}, t)] + \nonumber\\
&& \sum\limits_{\Delta \vec{\rho}} [1-m_1 (\vec{\rho}, t) m_1
(\vec{\rho}+ \Delta \vec{\rho}, t)]  \nonumber\\
&& \tanh \frac{J}{k_{\rm B}T} [m_2 (\vec{\rho}, t) + \frac{J_{\mathcal{S}}}{J}
\sum\limits_{\Delta \vec{\rho}\;'} m_1 (\vec{\rho}
+ \Delta \vec{\rho}\;', t) \nonumber\\
&& - m_2 (\vec{\rho} + \Delta \vec{\rho}, t)
 -\frac{J_{\mathcal{S}}}{J} \sum \limits_{\Delta \vec{\rho}\;'} m_1
(\vec{\rho} + \Delta \vec{\rho}\;' + \Delta \vec{\rho}, t)]. \quad
\end{eqnarray}
\end{itemize}
The equations for $n=D$ and $n=D+1$ are analogous.  The time $t$ in
Eqs.~(\ref{eq24},~\ref{eq25},~\ref{eq26}) is related to time $\tau$
in Eqs.~(\ref{eq6},~\ref{eq7},~\ref{eq8}) via
\begin{equation} \label{eq4p}
\tau=\frac{T_{\rm cb}/T-1}{8\tau_s\xi^2}.
\end{equation}
The numerical solutions of the set of equations
(\ref{eq24},~\ref{eq25},~\ref{eq26}) for quenching experiments from
infinite temperature to the states inside the miscibility gap is the
subject of the next section.

\section{Numerical Results for Phase Separation following Deep Quenches}
\begin{figure}
\includegraphics*[width=0.38\textwidth]{fig1a.eps}
\includegraphics*[width=0.33\textwidth]{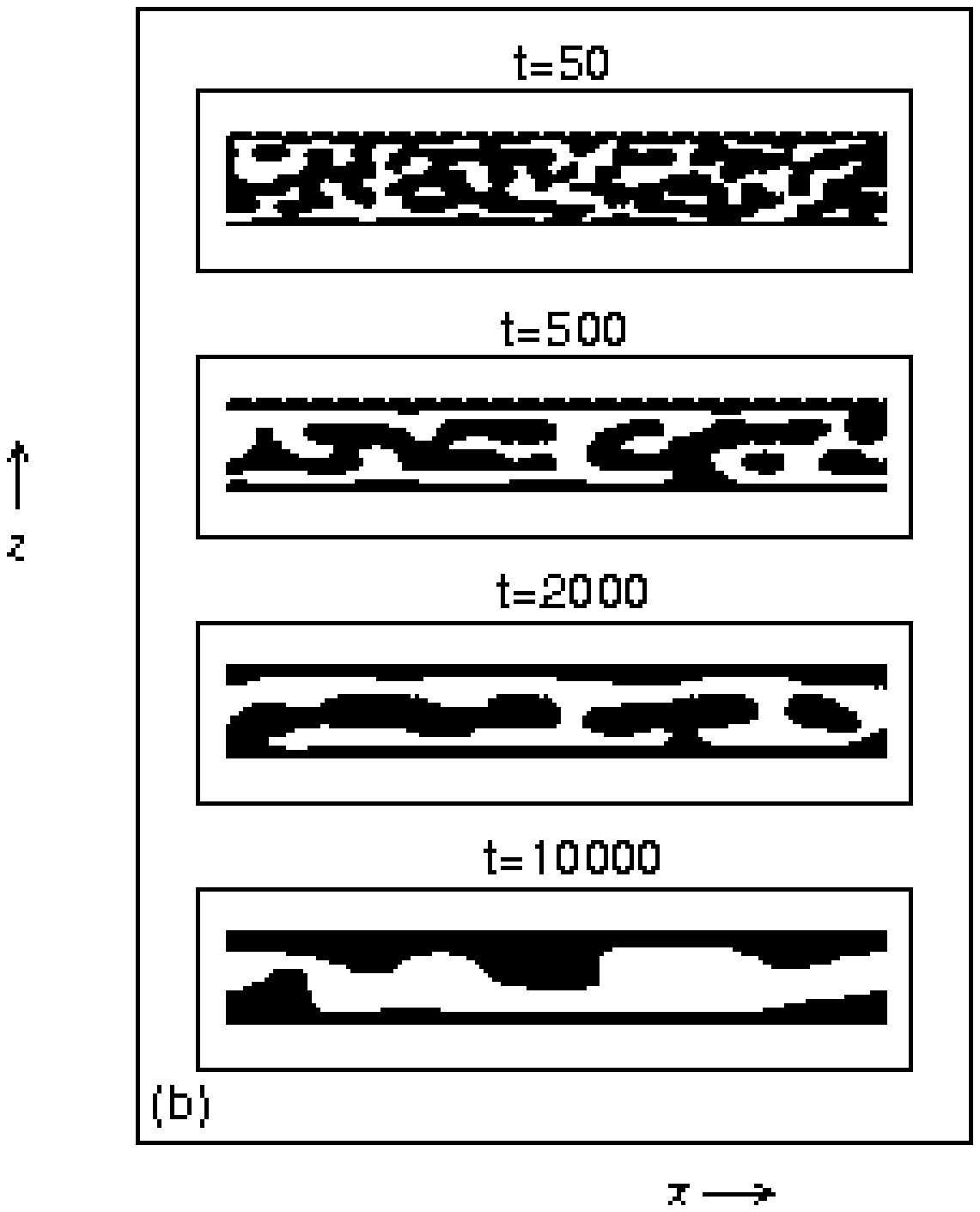}
\includegraphics*[width=0.33\textwidth]{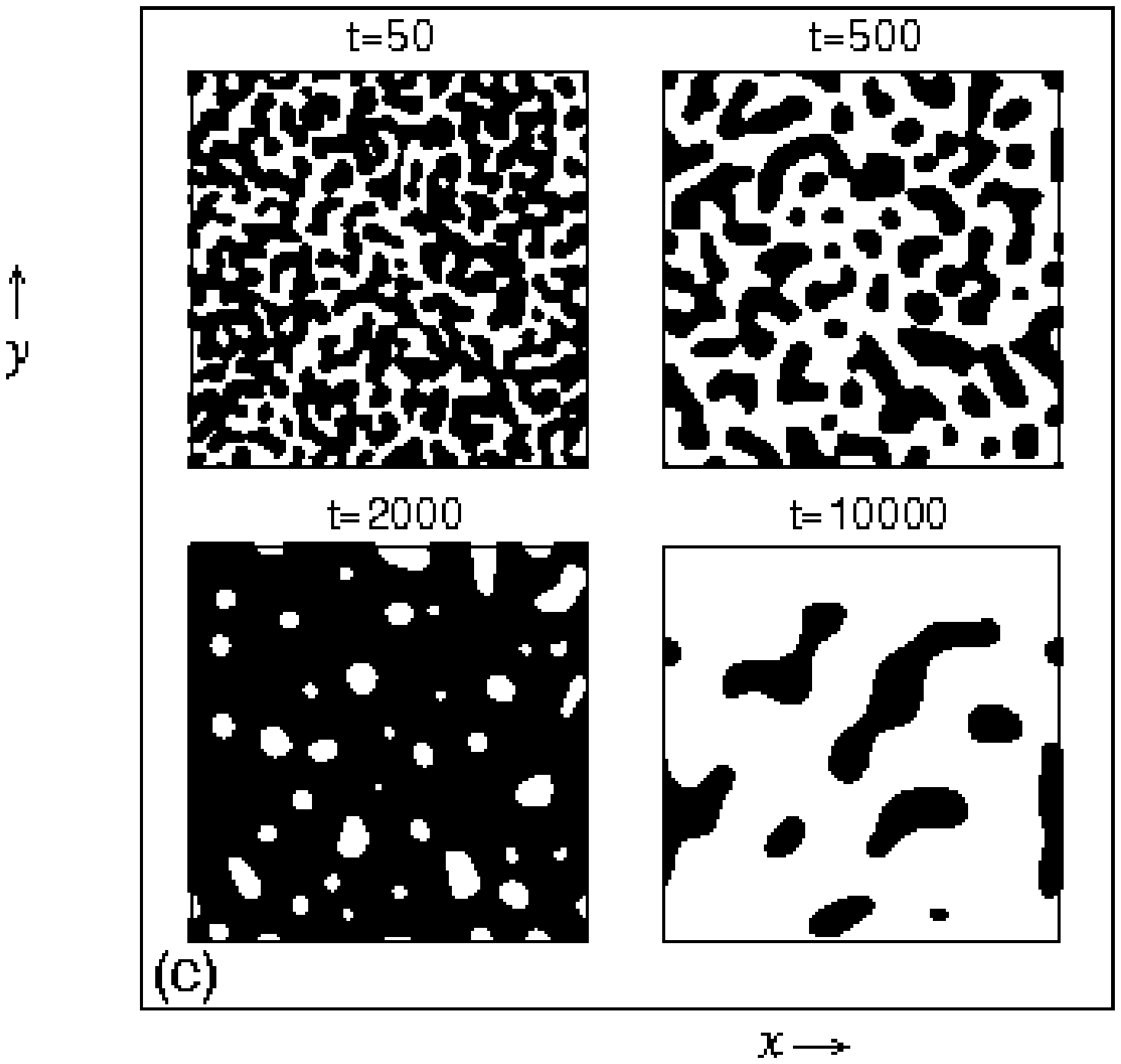}
\caption{\label{fig1} (a) Layerwise order parameter $\Psi_{av}(n)$
plotted vs. layer index $n$
for four different times for the choice 
$D=29, L=128$, and $H_{\mathcal{S}1}=H_{\mathcal{S}2}=1.0$.
The continuous lines are cubic interpolations to the original data,
used as guides to the eye.
(b) Cross-sectional snapshot pictures of the same systems as in (a),
displaying the magnetization configuration in the $xz$-plane for
$y=L/2$. If the magnetization at a lattice site is positive, a
black dot is printed.
(c) Same as (b), but for a plane parallel to the walls, at
$n=15$.}
\end{figure}

In this section we shall present selected results from the numerical
solutions of Eqs.~(\ref{eq24},~\ref{eq25},~\ref{eq26}), choosing three
values of the thickness of the film ($D= 9$, 19, and 29, corresponding
to $n_{\mbox{max}}= 10$, 20, and 30 lattice planes, respectively), and
a quench at time $t=0$ from infinite temperature to $T/T_{\rm cb}=2/3$,
for the special cases $H_{\mathcal{S}1}=H_{\mathcal{S}2}=1.0$, 0.1, and
$J_{\mathcal{S}}=J$. The initial conditions for $m_n(\vec{\rho},t)$ is
chosen by taking $m_n(\vec{\rho},0)$ from a random uniform distribution
between $-1$ and $+1$, with the total magnetization in the thin film zero.

\begin{figure}
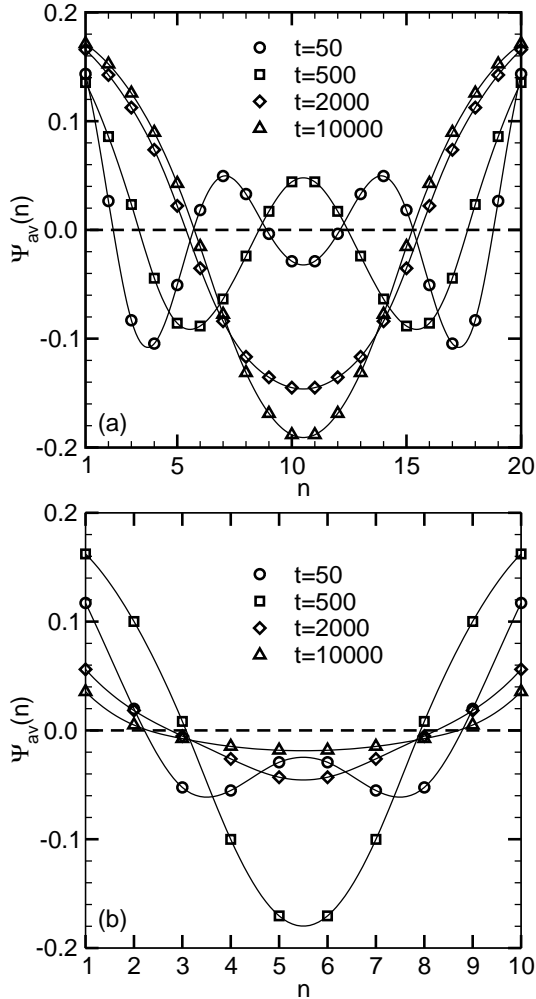

\includegraphics*[width=0.38\textwidth]{fig2a.eps}
\includegraphics*[width=0.38\textwidth]{fig2b.eps}
\caption{\label{fig2} Layerwise average order parameter $\Psi_{\rm av}(n)$ plotted vs. $n$
for four different times for the cases (a) $D=19, L=128$ and (b) $D=9,
L= 128$. Always $H_{\mathcal{S}1}=H_{\mathcal{S}2}=0.1$ is chosen.}
\end{figure}

At first sight a quench to $T/T_{\rm cb}=2/3$ does not look like a
particularly deep quench. However, one must keep in mind that in
order to have $\xi$ larger (or equal) than a lattice spacing one
must have $T \geq 5.57J/k_{\rm B}$, i.e. much closer to $T_{\rm
cb}=6J/k_{\rm B}$.  At the present temperature $k_{\rm B}T/J=4.0$,
the correlation length is as small as $\xi\simeq 0.33$ lattice
spacings. Following the standard reasoning for the simulation of the
Ginzburg-Landau model, Eqs.~(\ref{eq6},~\ref{eq7},~\ref{eq8}), one
should choose $2\xi$ as the size of the spatial discretization mesh:
dealing with the above physical film thickness would require rather
huge lattices. In addition, Eqs.~(\ref{eq1},~\ref{eq2},\ref{eq3})
do not represent the actual free energies of the lattice model
[Eqs.~(\ref{eq14},~\ref{eq15},~\ref{eq16},~\ref{eq17},~\ref{eq18})]
accurately at such low temperatures either.

\begin{figure}
\includegraphics*[width=0.33\textwidth]{fig3a.eps}
\includegraphics*[width=0.33\textwidth]{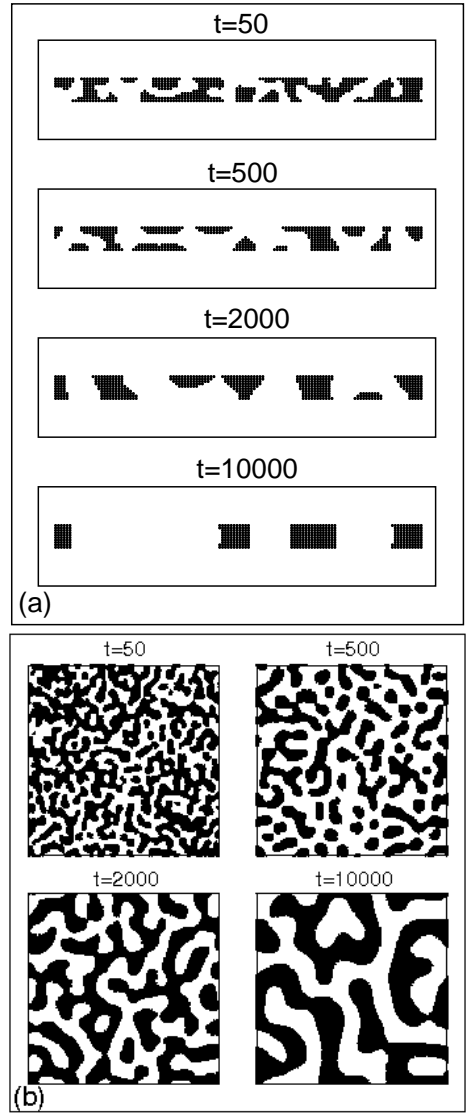}
\caption{\label{fig3} Cross-sectional snapshot pictures of the system with
$D=9, L=128, H_{\mathcal{S}1}=H_{\mathcal{S}2}=0.1$ for (a) the $xz$-plane and 
(b) the $xy$-plane at $n=5$ for the same system
as in Fig. \ref{fig2}(b).}
\end{figure}

Choosing the time unit $\tau_s=1$ in
Eqs.~(\ref{eq24},~\ref{eq25},~\ref{eq26}) we have found that accurate
numerical solutions of these equations result already when one chooses
a rather large discrete time step, $\delta t = 0.1$. Other than this
discretization of time no approximations whatsoever enter the numerical
solution. Of course, one always has to deal with a finite system
geometry also in lateral directions: unless otherwise mentioned we
choose $L_x=L_y=L=128$ for all values of $D$ and apply periodic boundary
conditions in both the directions.

\begin{figure}
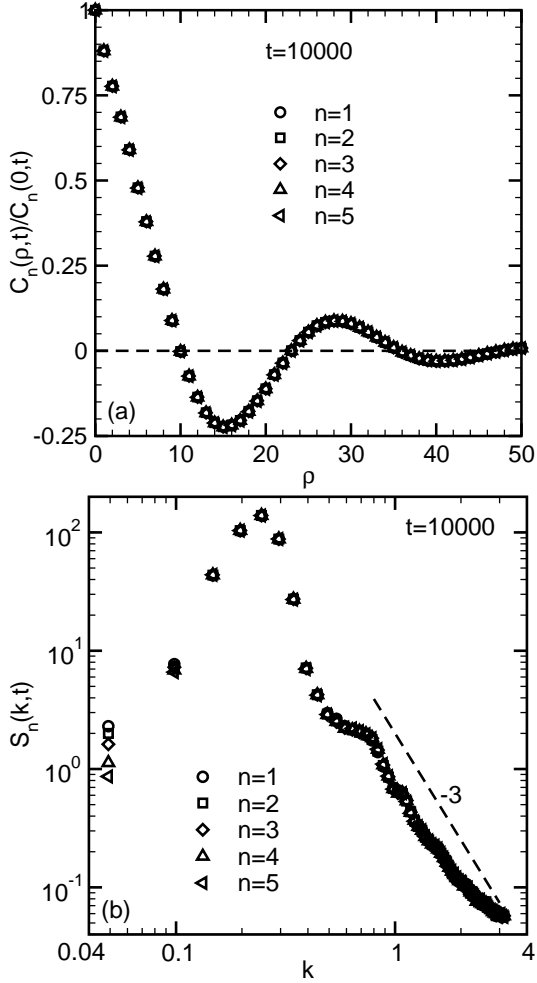

\includegraphics*[width=0.38\textwidth]{fig4a.eps}
\includegraphics*[width=0.38\textwidth]{fig4b.eps}
\caption{\label{fig4} (a) Layerwise correlation function $C_n(\rho,t)$
plotted vs.
$\rho$, for the choice $H_{\mathcal{S}1}=H_{\mathcal{S}2}=0.1$, $D=9,L=128, t=10000.$ Data for
$n=1,2,3,4$ and 5 superimpose almost exactly. (b) Fourier
transform $S_n(k,t)$ of $C_n(\rho,t)$, again resolved with respect to
individual layers. In (b) the dashed line corresponds to the
Porod tail $k^{-3}$.}
\end{figure}

Figure~\ref{fig1} shows data for a typical time evolution for $D=29$
and $H_{\mathcal{S}1}=H_{\mathcal{S}2}=1.0$. In Fig.~\ref{fig1}(a), we
show the layerwise average order parameter, $\Psi_{\rm av}(n)=L^{-2}
\sum \limits _{\vec{\rho}}m_n(\vec{\rho},t) $, as a function
of the layer index $n$.  One sees that in the surface planes the
magnetization takes its saturation value rather fast, as expected due
to the large surface fields. Then the magnetization decreases very
rapidly, already for $n=3$ the magnetization for $t=50$ is strongly
negative. For $t=50$ the curve $\Psi_{\rm av}(n)$ then exhibits the
oscillations typical for ``surface-directed spinodal decomposition''
\cite{7,8,9,10,11,12,13,14,15,16,17,18,19}, with a second maximum at
$n=7$ and a (weak) third one at $n=13$.  Because of the symmetric surface
fields, it is expected that the profile should be symmetric around the
center of the film,
\begin{equation}\label{eq27}
\Psi_{\rm av}(n)=\Psi_{\rm av}(D-n+2).\quad 
\end{equation}
But in reality this is not obeyed because of finite system size and
lack of averaging over sufficiently large number of independent initial
configurations (in our case averaging was done only over $5$ independent
random initial configurations). However, all plots for $\Psi_{\rm
av}(n)$ have been symmetrized by hand by taking advantage of property
(\ref{eq27}).

\begin{figure}
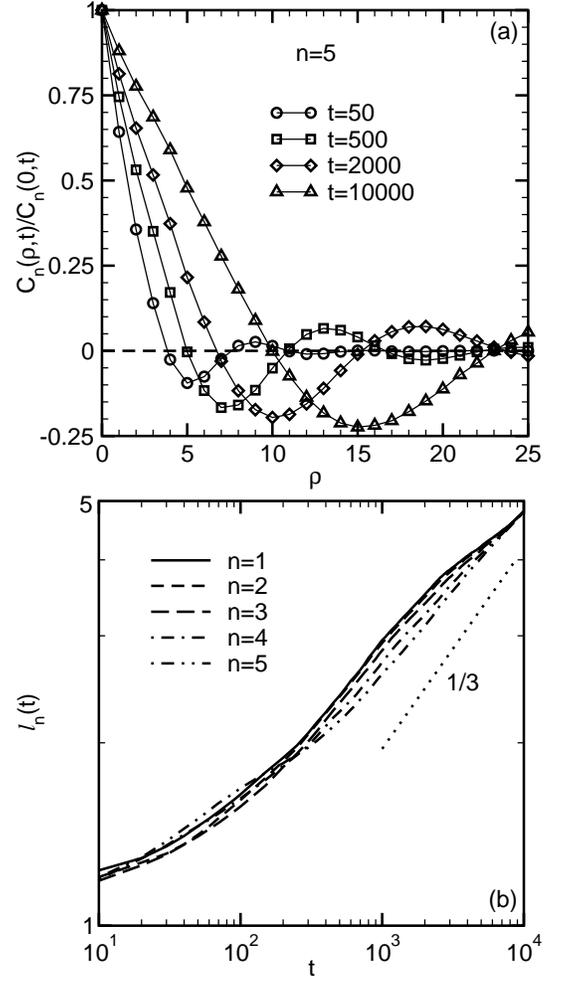

\includegraphics*[width=0.38\textwidth]{fig5a.eps}
\includegraphics*[width=0.38\textwidth]{fig5b.eps}
\caption{\label{fig5} (a) Same as Fig.~\ref{fig4}(a) but only for $n=5$, and four
times as indicated. (b) Characteristic domain length $\ell_n(t)$ of individual layers,
shown on a
log-log plot versus time $t$. The dotted line corresponds to
LS growth law $t^{1/3}$.}
\end{figure}

One can further see from Fig.~\ref{fig1}(a) that the thickness of the
surface enrichment layers at the walls slowly grows with increasing
time, which is also obvious from Fig.~\ref{fig1}(b) where we present the
snapshot pictures of vertical sections ($xz$ plane).  At a later time the
position of the second peak of $\Psi_{\rm av}(n)$ has moved towards the
center (it now occurs at $n=11$ for $t=500$) with a pronounced minimum
in the center of the film.  However, for $t \geq 2000$ this second peak
has merged with its mirror image, i.e., now in the film center there is a
maximum of $\Psi_{\rm av}(n)$ rather than a minimum. At very late time,
this central maximum has disappeared again ($t=10 000)$ and the profile
looks like that of a simple stratified structure. Actually this is not
the case, as seen in Fig.~\ref{fig1}(c), where we have shown the snapshot
pictures parallel to the surfaces ($xy$ plane) for $n=15$.  As is also
observed in the Ginzburg-Landau studies of surface-directed spinodal
decomposition \cite{8,9,10,12,13,14,17,18}, in the lateral directions
one can observe initially a rather random pattern which rapidly coarsens
with increasing time.

\begin{figure}
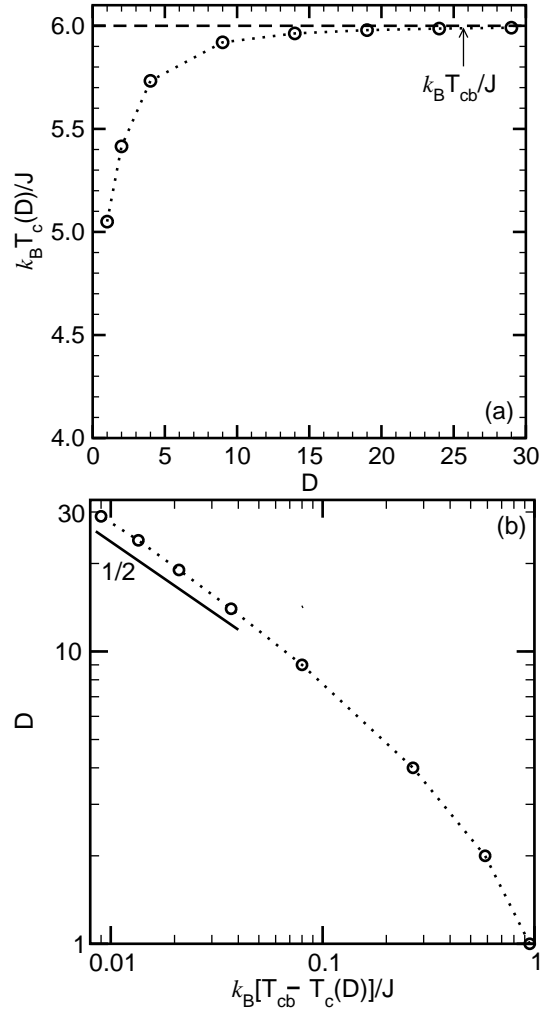

\includegraphics*[width=0.38\textwidth]{fig6a.eps}
\includegraphics*[width=0.38\textwidth]{fig6b.eps}
\caption{\label{fig6} Plot of finite-size critical temperature $T_c(D)$ 
versus film thickness $D$, (a) on a linear scale and (b) on
a log-log plot to display the asymptotic behavior $T_{\rm cb}
- T_{\rm c}(D) \propto D^{-2}$ of mean field theory. The dashed line in (a)
marks the bulk critical temperature, whereas the continuous line in (b)
corresponds to the asymptotic theoretical prediction.}
\end{figure}

Figure~\ref{fig2} shows other examples where we have chosen thinner
films (viz., $D=9$ and $D=19$) and a much weaker boundary field
$(H_{\mathcal{S}1}=H_{\mathcal{S}2}=0.1)$.  Now the amplitude of the
variation of $\Psi_{\rm av}(n)$ is much smaller, and at late times the
order parameter profiles across the film have almost no structure. The
explanation for this behavior is seen in Fig.~\ref{fig3} where we
show snapshot pictures of the states evolving for the choice $D=9$
of Fig.~\ref{fig2}(b): The system develops towards a two-dimensional
arrangement of columns of positive magnetization connecting the two walls,
and thus in each plane ($n = \textrm{const}$), there is only a weak excess
of magnetization in any one direction.  These results qualitatively do
not differ from the numerical studies of spinodal decomposition in thin
films based on the Ginzburg-Landau equation, such as \cite{18}; however,
the advantage of the present treatment is that the parameters of the model
have an immediate and straightforward physical meaning.  The fact that
in the late stages there is almost no nontrivial structure across the
film is also evident from a comparison of the pair correlation function
$C_n(\rho,t)$ [$=\langle m_n(\vec 0,t)m_n(\vec\rho ,t)\rangle - \langle
m_n(\vec 0,t)\rangle \langle m_n(\vec\rho,t)\rangle $] and their Fourier
transforms $S_n(k,t)$ in different layers, shown in Fig.~\ref{fig4}.

\begin{figure}
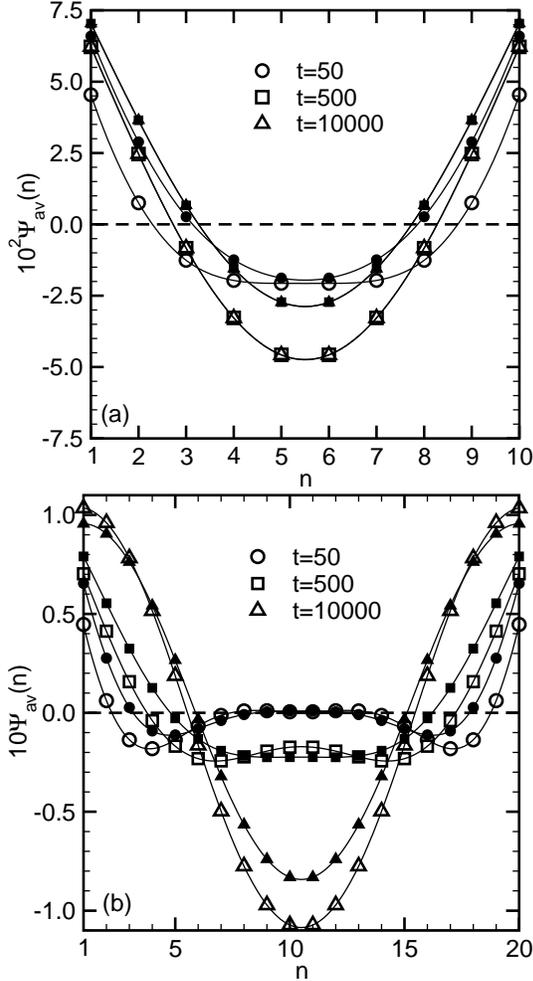

\includegraphics*[width=0.38\textwidth]{fig7a.eps}
\includegraphics*[width=0.38\textwidth]{fig7b.eps}
\caption{\label{fig7} Comparison of the layerwise average order parameter profiles
across the film for three times $(t=50, 500$ and 10000,
respectively), according to the lattice model (open symbols) and
the GL model (filled symbols), for (a) $D=9$, and (b) $D=19$,
respectively. All data refer to the choice $H_{\mathcal{S}1}=H_{\mathcal{S}2}=0.1,\;
k_{\rm B}T/J=5.875$.}
\end{figure}

In Fig.~\ref{fig5}(a), the plot of the time evolution of $C_n(\rho,t)$,
for $n=5$, for the same system as in Fig.~\ref{fig3}, clearly reflects
the coarsening behavior. Note that the apparent nonscaling behavior of
$C_n(\rho,t)$ is due to strong fluctuation of the layerwise average order
parameter at early time. In Fig.~\ref{fig5}(b) we plot the layerwise
average domain size $\ell_n(t)$ as a function of $t$ extracted from the
condition $C_n(\rho=\ell_n(t),t)=C_n(0,t)/2$.  The characteristic length
$\ell_n(t)$ initially grows rather slowly, for times $10 < t < 1000$ there
is considerable curvature on the log-log plot, while for $t \geq 1000$ the
behavior is already close to the standard Lifshitz-Slyozov (LS) \cite{43}
$\ell (t) \propto t^{1/3}$ law. The fact that the ``effective exponent''
$d[\ln \ell(t)]/d (\ln t)$ approaches 1/3 from below is quite reminiscent
of Monte Carlo simulations of coarsening in the two-dimensional Kawasaki
spin-exchange model \cite{44}, of course.  It should be noted that Monte
Carlo simulations include thermal fluctuations that are absent in our
molecular field treatment. However, it is generally believed \cite{4}
that thermal fluctuations are irrelevant during the late stages of
coarsening. In view of these facts, the similarity of our results with
the previous Monte Carlo studies of coarsening on lattice models is not
unexpected. The advantage of the present approach in comparison with
Monte Carlo, however, is that much less numerical effort is needed.

\section{The Critical Region: A Comparison between the Lattice
Approach and the Ginzburg-Landau treatment}

Near the critical point one must take into account that the critical point
is slightly shifted to lower temperatures for films of finite thickness
as compared to the bulk \cite{21,31,32,33,34}. In Fig.~\ref{fig6}(a),
we plot the critical temperatures, $T_{\rm c}(D)$, for films of finite
width as a function of width $D$. The numerical values for $T_{\rm c}(D)$
were obtained by solving Eqs.~(\ref{eq10},~\ref{eq11},~\ref{eq12}) for
$H=H_{\mathcal{S}1}=H_{\mathcal{S}2}=0$ starting  with the assignment of
uniform magnetization to the layers in a random fashion so that the total
film magnetization is zero. Note that as $D\rightarrow\infty$, $T_{\rm
c}(D)$ is expected to approach its $3$-$d$ mean field value $6J/k_{\rm
B}=T_{\rm cb}$ and in the limit $D\rightarrow 0$, we expect the $2$-$d$
value $4J/k_{\rm B}$. In Fig.~\ref{fig6}(b), we present the deviation
of $T_{\rm c}(D)$ from $T_{\rm cb}$ as a function of $D$ on a log-log
plot. From finite-size scaling theory, this difference should vanish as
$T_{\rm cb}-T_{\rm c}(D)\sim D^{-2}$.

While for $D=29$ we have $T_{\rm c}(D) \simeq 5.99 J/k_{\rm B}$, for
$D=9$ we have $T_{\rm c}(D)\simeq 5.92 J/k_{\rm B}$. Since lateral
phase separation occurs only for $T <T_{\rm c}(D)$ \cite{35,36,37},
of course, this shift restricts the range of $\xi$ that can be studied
for the present choices of $D$. While $2\xi \simeq 1.0 $ occurs for
$T=4.75 J/k_{\rm B}$ and hence this choice, for which the cell size
of the Ginzburg Landau model agrees with the lattice spacing of the
molecular field model, is safely accessible, already for $\xi \simeq 2.0$
(occurring for $k_{\rm B}T=5.875J$) we are only slightly below $T_{\rm
c}(D=9)$, and for $\xi = 3$ we would already be in the one-phase region of
such a thin film.  In view of these considerations, $k_{\rm B}T=5.875J$
was chosen as a temperature where it makes sense to compare the lattice
model with $D=9$, 19, and 29 (i.e., $n_{\mbox{max}}=10$, 20, and 30
lattice planes, respectively) with the corresponding Ginzburg Landau
(GL) model.  Note that for the convenience of comparison as well as
accuracy of numerical solutions, the discrete mesh sizes $\Delta x$,
$\Delta y$, and $\Delta z$ in the GL model were adjusted to the lattice
constant $a$ and we have set $\tau_0=0$.

\begin{figure}
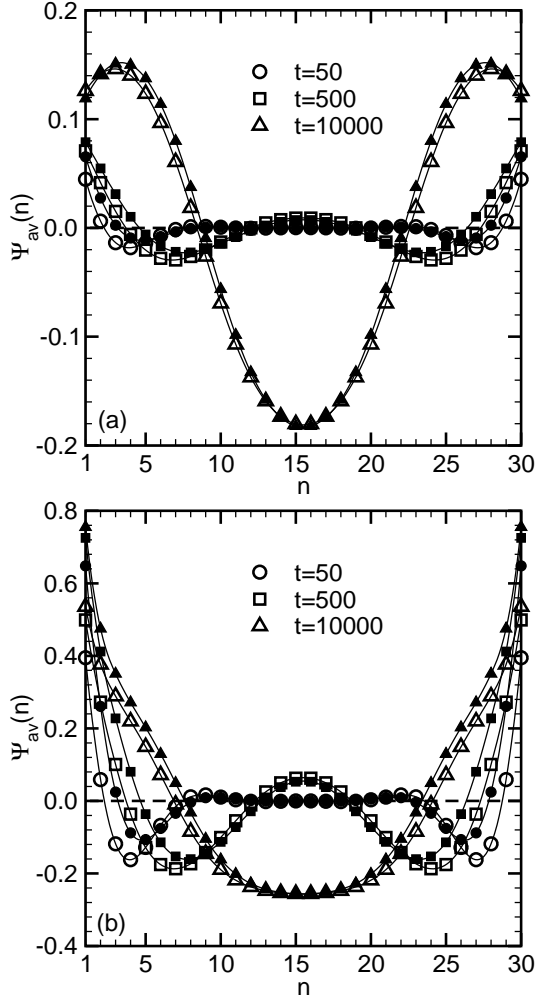

\includegraphics*[width=0.38\textwidth]{fig8a.eps}
\includegraphics*[width=0.38\textwidth]{fig8b.eps}
\caption{\label{fig8} Same as Fig.~\ref{fig7}, but
for $D=29$, and two choices of the strength of the surface fields,
$H_{\mathcal{S}1}=H_{\mathcal{S}2}=0.1$ (a) and $1.0$ (b).}
\end{figure}

\begin{figure}
\includegraphics*[width=0.33\textwidth]{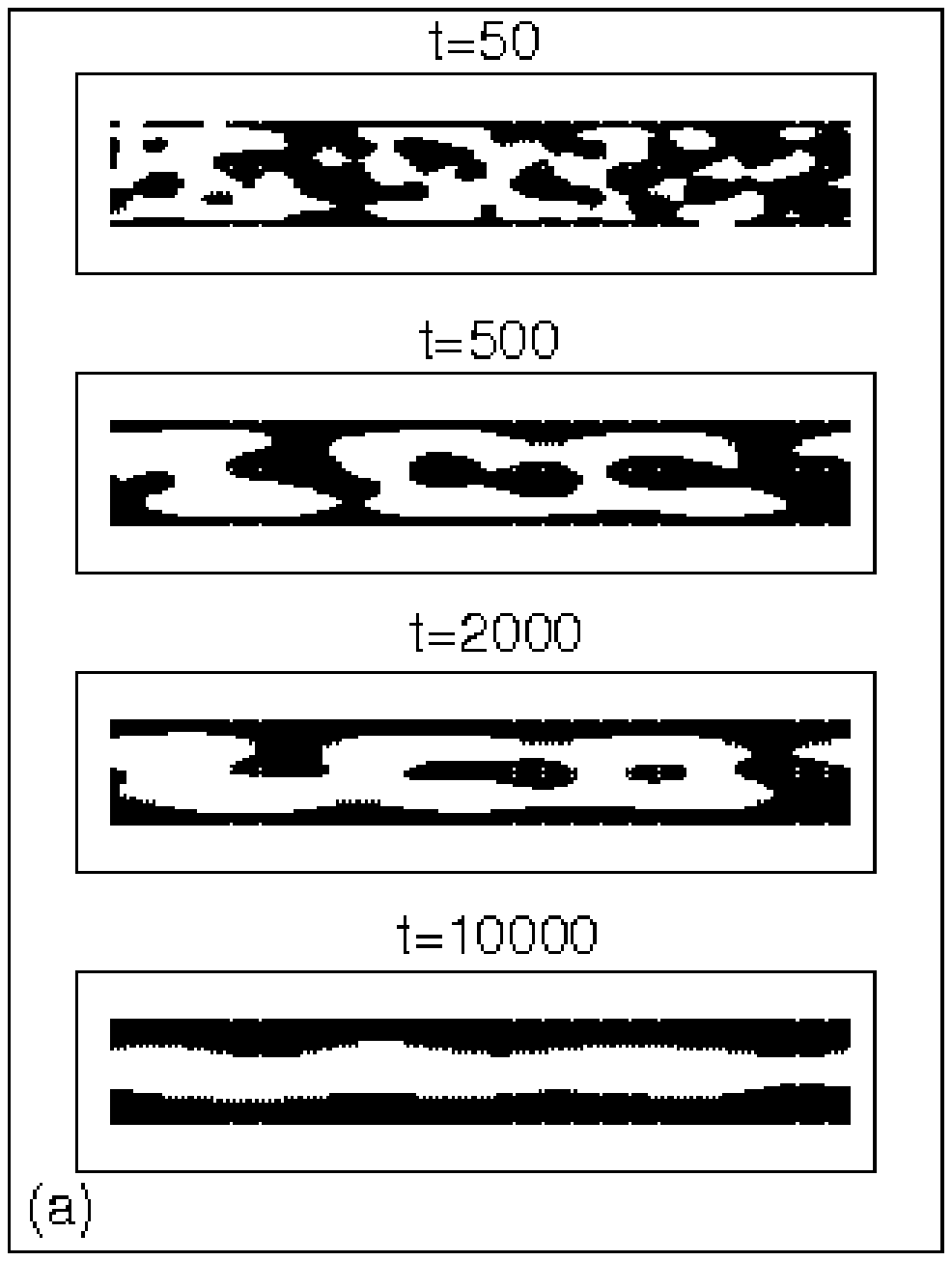}
\includegraphics*[width=0.33\textwidth]{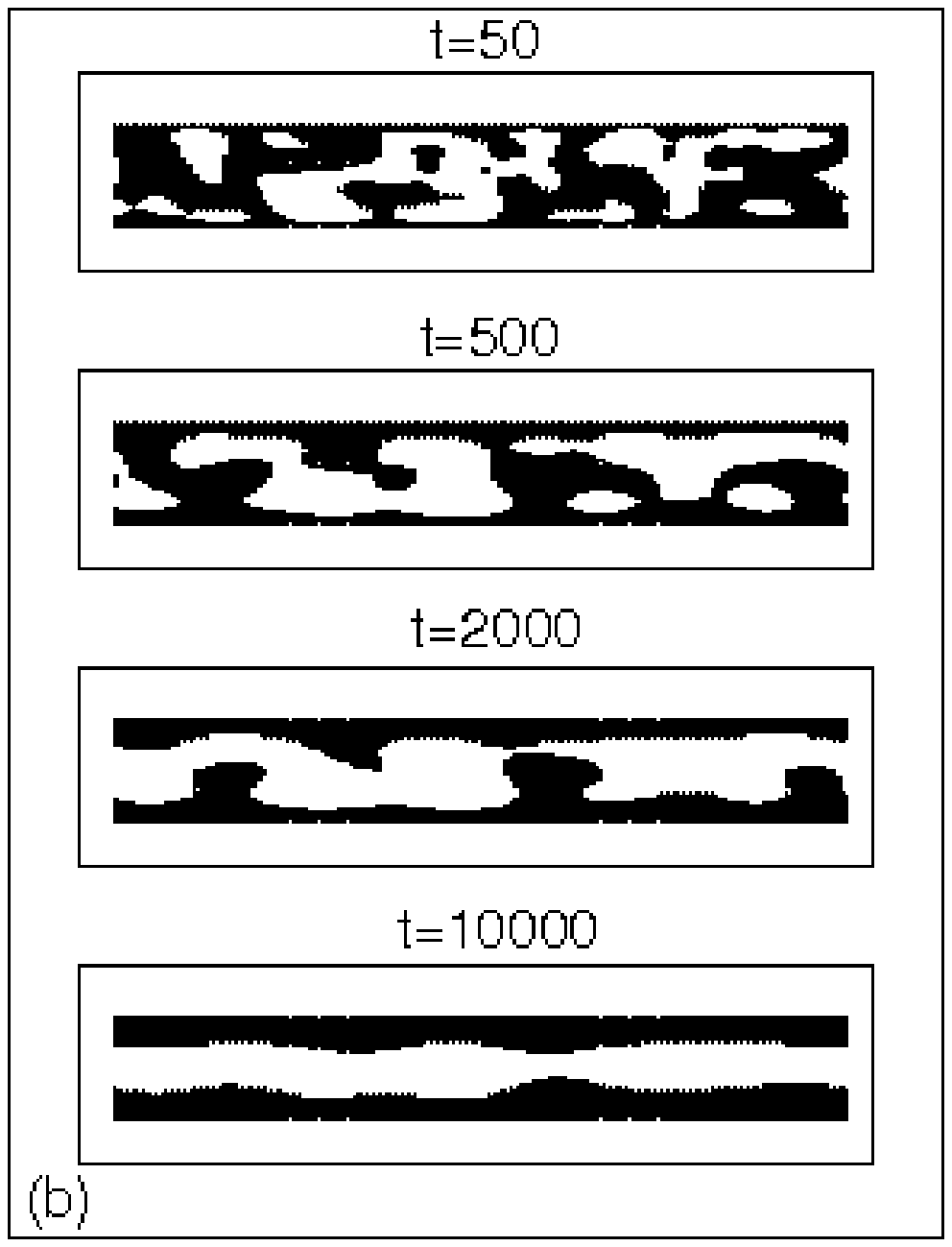}
\caption{\label{fig9} Cross-sectional snapshot
pictures in the $xz$-plane, for $D=29, H_{\mathcal{S}1}=H_{\mathcal{S}2}=0.1$, at $k_{\rm B}T/J=5.875$
[same system as in Fig~\ref{fig8}(a)],
for four different times as indicated, according to (a) the lattice model
and (b) the Ginzburg Landau model.}
\end{figure}

\begin{figure}
\includegraphics*[width=0.36\textwidth]{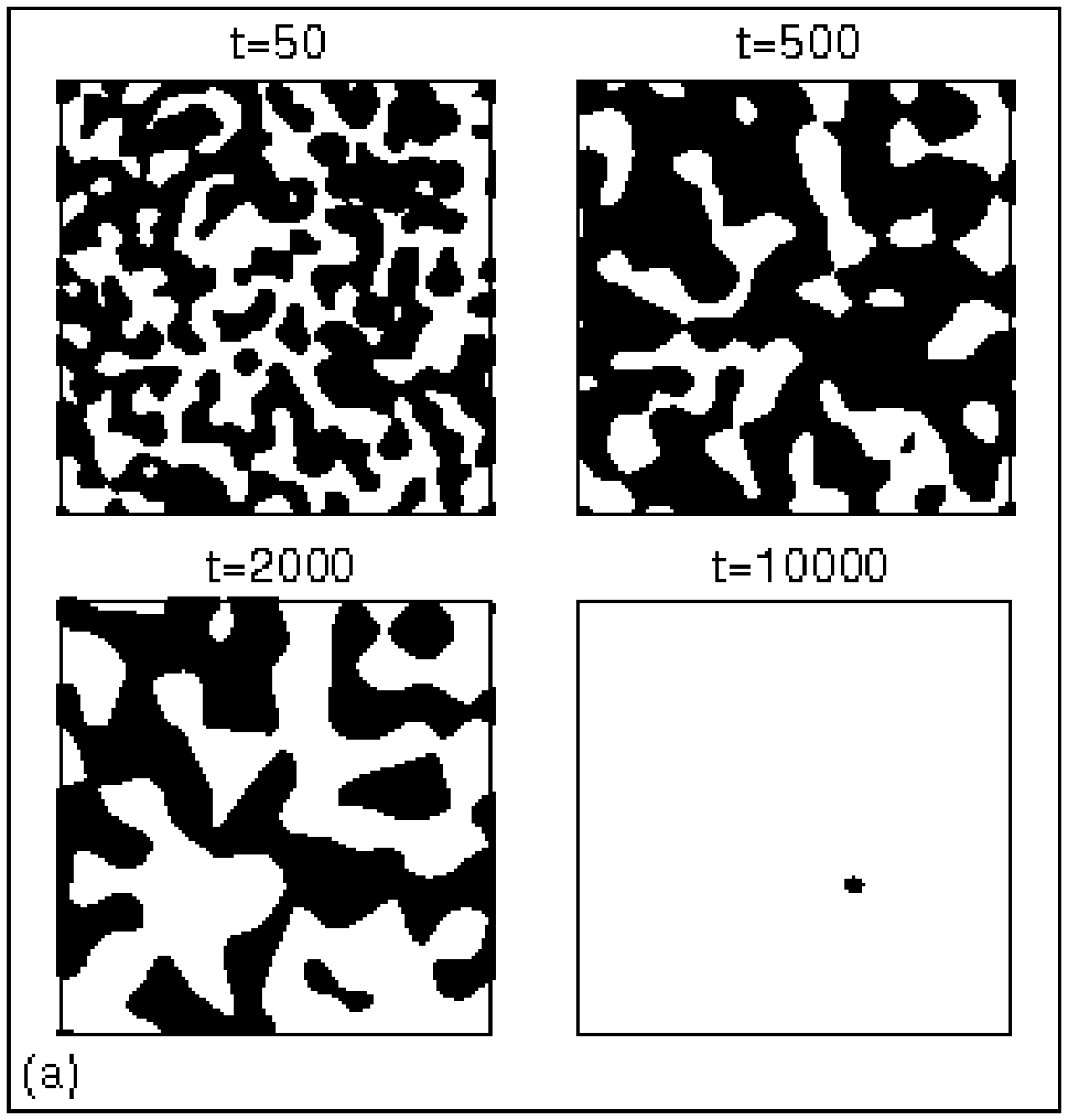}
\includegraphics*[width=0.36\textwidth]{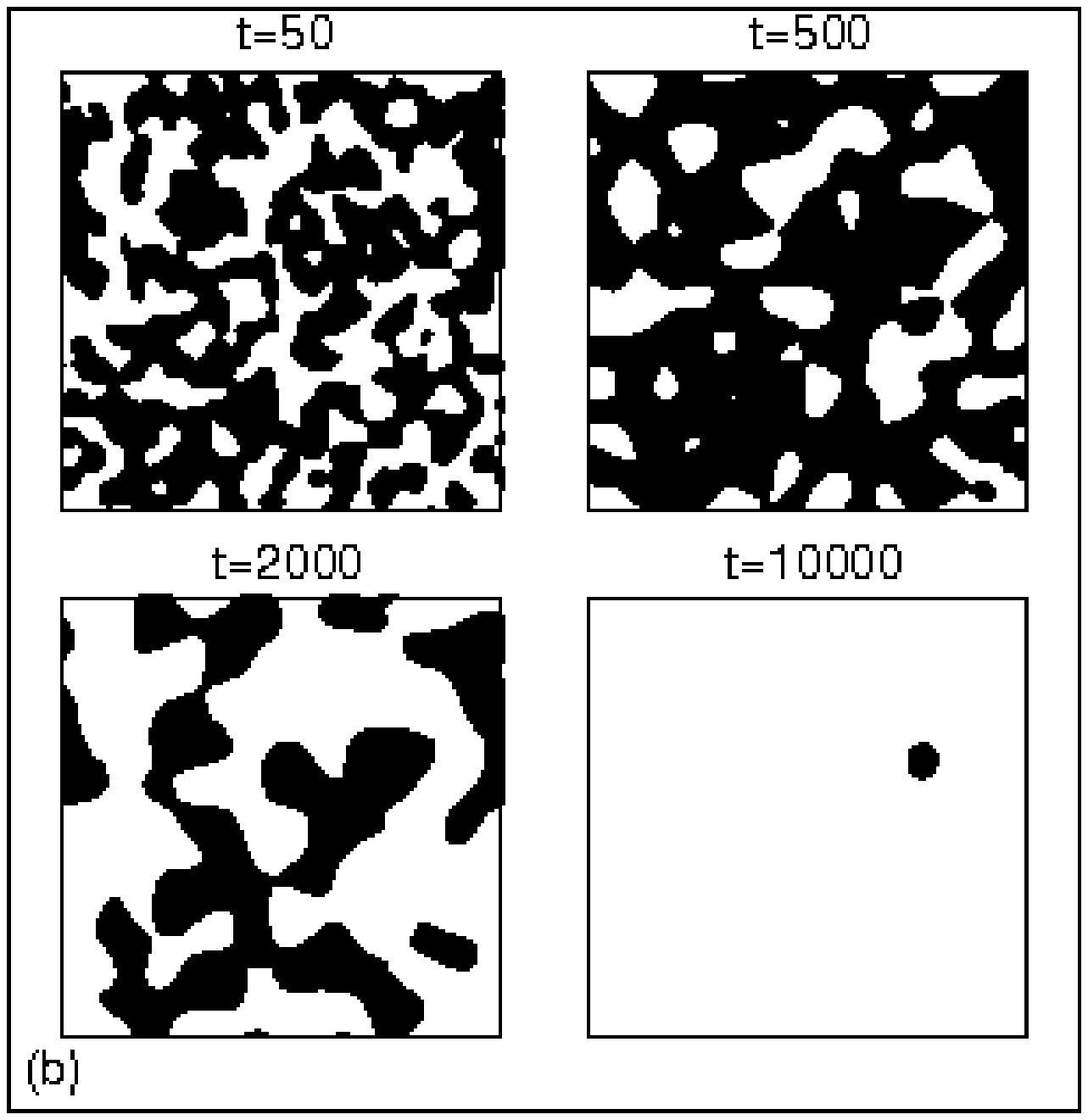}
\caption{\label{fig10} Same as Fig.~\ref{fig9},
but for a plane at $n=15$, parallel to the walls.}
\end{figure}

\begin{figure}
\includegraphics*[width=0.36\textwidth]{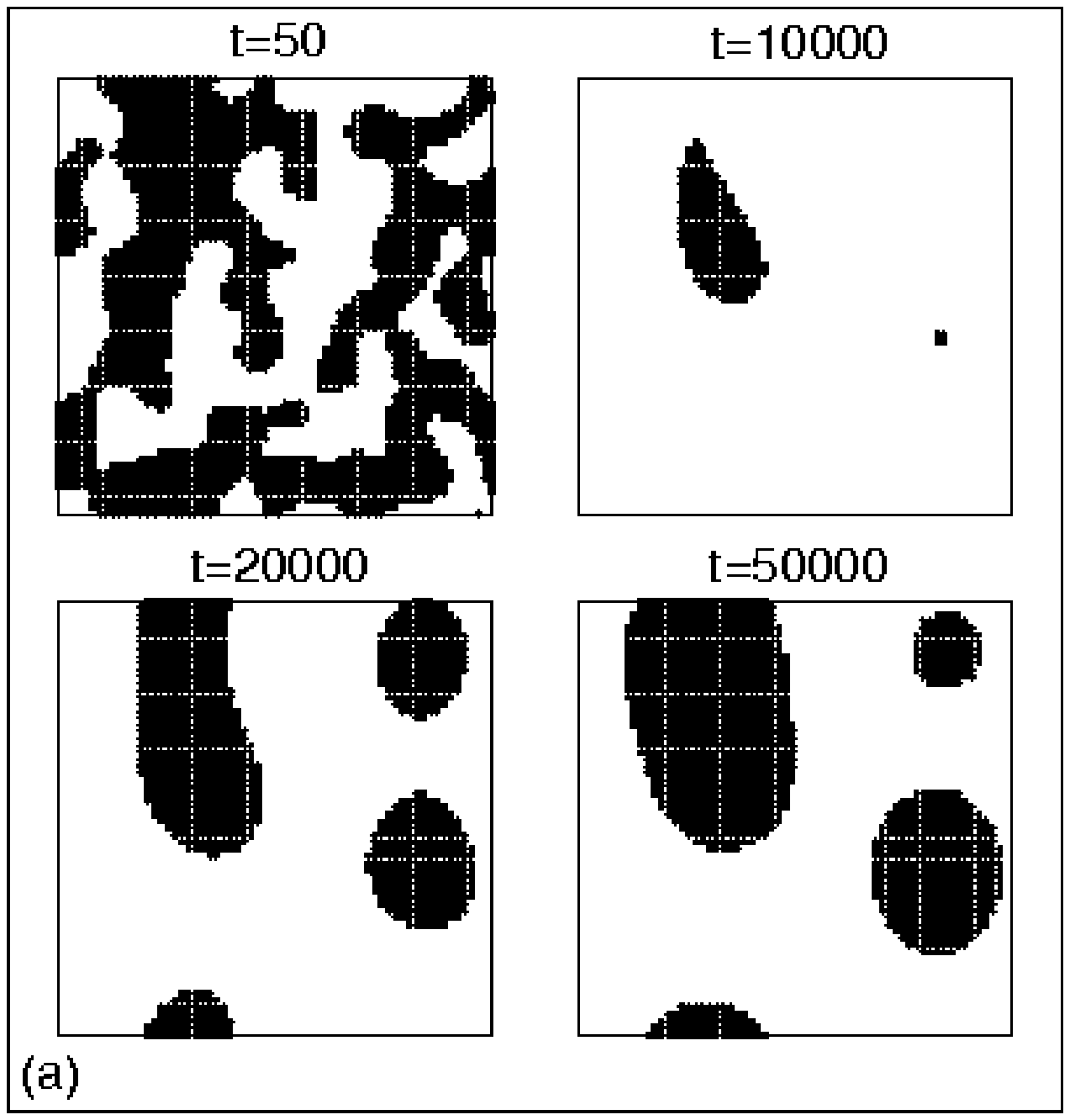}
\includegraphics*[width=0.36\textwidth]{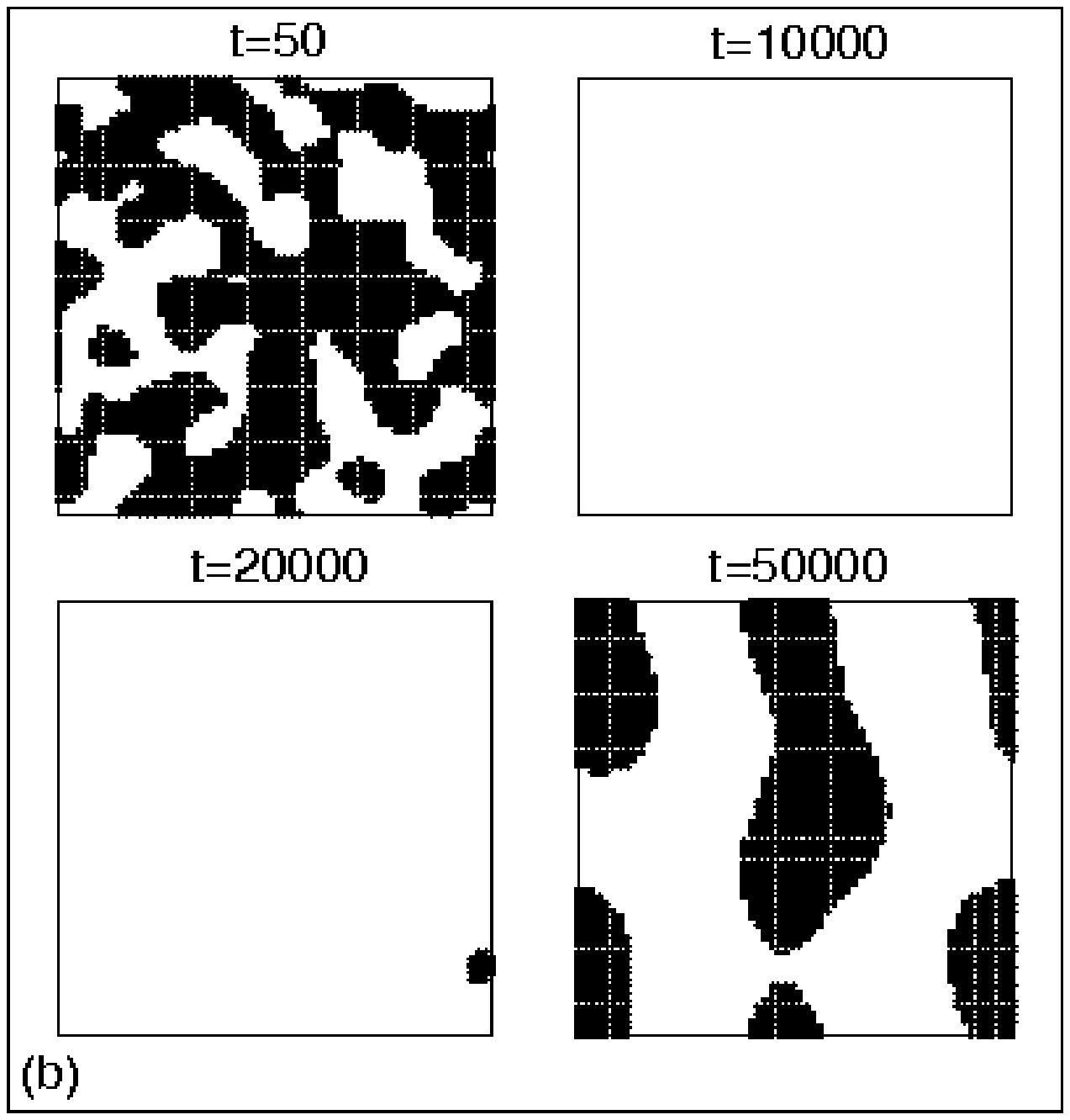}
\caption{\label{fig11} Same as Fig.~\ref{fig10},
but for $D=19,~L=64$, at $n=10$.}
\end{figure}

In Fig.~\ref{fig7} we present the comparison between the lattice model and
GL model at $k_{\rm B}T=5.875J$ ($\xi\simeq 2.0$, $m_{\rm b}\simeq 0.25$)
with $H_{\mathcal{S}1}=H_{\mathcal{S}2}=0.1$ for $D=9$ and $19$. At this
temperature the parameters $h_{\mathcal{S}1}$, $g$, $\gamma$, $\tau$
have the values $52.3$ ($523H_{\mathcal{S}1}$), -128, 32, and $t/1600$,
respectively.  One sees that for $D = 9$ [Fig.~\ref{fig7}(a)] the behavior
of the lattice model and the GL model are qualitatively similar, but there
is no quantitative agreement. These discrepancies do get smaller, however,
with increasing film thickness [Fig.~\ref{fig7}(b)]. In Fig.~\ref{fig8},
where we plot the order parameter profile for D = 29 corresponding to
two values of surface fields (see caption for details), we see that
the discrepancies between the lattice and GL models are already quite
small, irrespective of the choice of the surface field.  However,
for stronger surface field [Fig.~\ref{fig8}(b)], the discrepancy is
still visible at the surfaces. We expect that the comparison should
be perfect for any surface field in the semi-infinite limit. In the
immediate vicinity of the critical point, the close correspondence
between the time evolution predicted by the lattice theory and the GL
model is also evident when one compares snapshot pictures of the time
evolution, shown in Figs.~\ref{fig9}, \ref{fig10}, generated at $k_{\rm
B}T/J=5.875$ corresponding to $H_{\mathcal{S}1}=H_{\mathcal{S}2}=0.1$,
$D=29$, for both the models.  However these snapshots suggest that at
this temperature the lateral inhomogeneity disappears in the late stages
of the phase separation process. Note that due to the proximity of the
critical point, which for finite $D$ is shifted and nonzero surface
fields, the coexistence curve separating the two-phase region from the
one-phase region may be considerably distorted \cite{29,31}.  In view of
that, it is plausible that the state point for small $D$ falls in the
one-phase region of the thin film. However, this did not happen in the
present case, as is clear from Fig.~\ref{fig11}, where we have shown
snapshot pictures over a much longer time scale for a system with all
parameters same as in Figs.~\ref{fig9}, \ref{fig10} except now we have
$D=19,~L=64$. So, the apparent stratified structure in Figs.~\ref{fig9},
\ref{fig10} is temporary which disappears at later time.


\begin{figure}
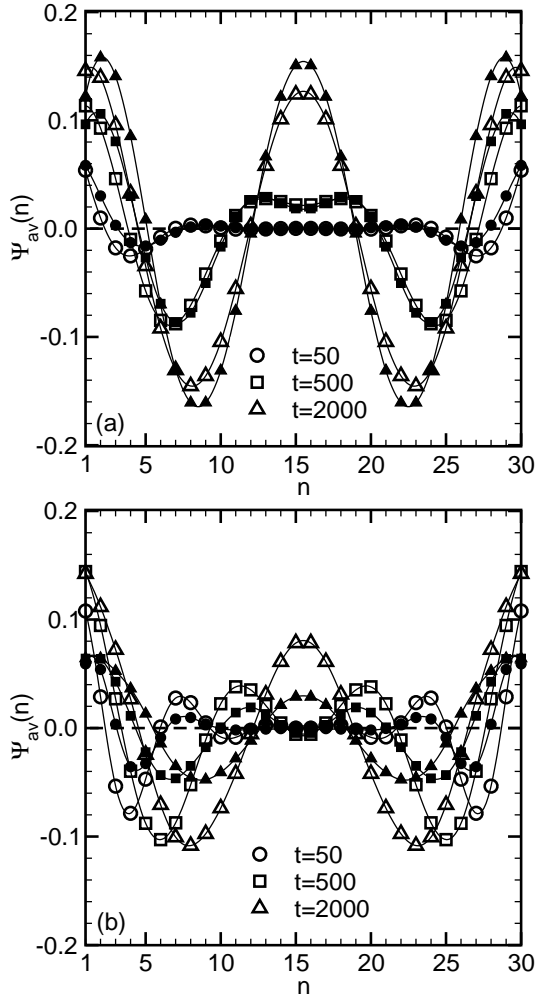

\includegraphics*[width=0.38\textwidth]{fig12a.eps}
\includegraphics*[width=0.38\textwidth]{fig12b.eps}
\caption{\label{fig12} Plot of $\Psi_{\rm av}(n)$ vs. $n$ for $D=29$
with $H_{\mathcal{S}1}=H_{\mathcal{S}2}=0.1$ at temperatures (a) $k_{\rm
B}T=5.57J$ and (b) $k_{\rm B}T=4.75J$.  The open symbols corresponds to
the lattice model whereas the filled symbols are for the GL model. Note
that $k_{\rm B}T=5.57J$ corresponds to bulk correlation length $\xi\simeq
1.0$ and at $k_{\rm B}T=4.75J$, $\xi\simeq 0.5$.}
\end{figure}

In Fig.~\ref{fig12} we show comparisons at lower temperatures,
viz., $k_{\rm B}T=5.57J$ ($\xi\simeq 1.0$, $m_{\rm b}\simeq 0.45$,
$h_{\mathcal{S}1}\simeq 15.4H_{\mathcal{S}1}$, $g=-8$, $\gamma=4$,
$\tau\simeq t/104$) and $k_{\rm B}T\simeq 4.75J$ ($\xi\simeq 0.5$,
$m_{\rm b}\simeq 0.72$, $h_{\mathcal{S}1}\simeq 0.44H_{\mathcal{S}1}$,
$g=-0.5$, $\gamma=0.5$, $\tau\simeq t/7.6$).  For both the temperatures
we have set $H_{\mathcal{S}1}=H_{\mathcal{S}2}=0.1$.  Rather pronounced
discrepancies between the lattice model and the GL model do occur,
however, at low temperature [Fig.~\ref{fig12}(b)], as expected.

Note that the prefactor $\gamma$ [see Eq.~(\ref{eq4})] in the scaled
Eq.~(\ref{eq7}) is an approximation which applies in the close vicinity of
the critical point \cite{8}. Here we try to take into account correction
terms to the leading behavior of Eq.~(\ref{eq7}) to make the GL model
more accurate for temperatures away from criticality. The order parameter
$\phi(\vec{\rho}, z,t)$ (not normalized by $m_{\rm b}$, and lengths not
rescaled by 2$\xi$) satisfies the boundary conditions \cite{8}
\begin{eqnarray}\label{eq28}
2 \tau_s \frac{\partial \phi (\vec{\rho},z=0,t)}{\partial t} &=&
\frac {H_1}{T} + \frac J T \left( 4 \frac {J_s}{J} -5 \right) \phi
(\vec{\rho},z=0,t)~~ \nonumber \\
& - & \left( \frac{T_{\rm cb}}{T} - 1 - \frac JT \right) 
\frac{\partial \phi (\vec{\rho},z,t)}{\partial z} |_{z=0}, 
\end{eqnarray}
\begin{eqnarray}\label{eq29}
\frac {\partial}{\partial z} \{(\frac {T_{\rm cb}}{T} -1) \phi (\rho ,z
,t)-\frac 1 3 [\phi (\rho, z ,t)]^3  \nonumber \\
 + \frac J T \frac {\partial ^2}{\partial z^2} [\phi(\rho,z,t)]\}
 | _{z=0} = 0.\quad
\end{eqnarray}

\begin{figure}
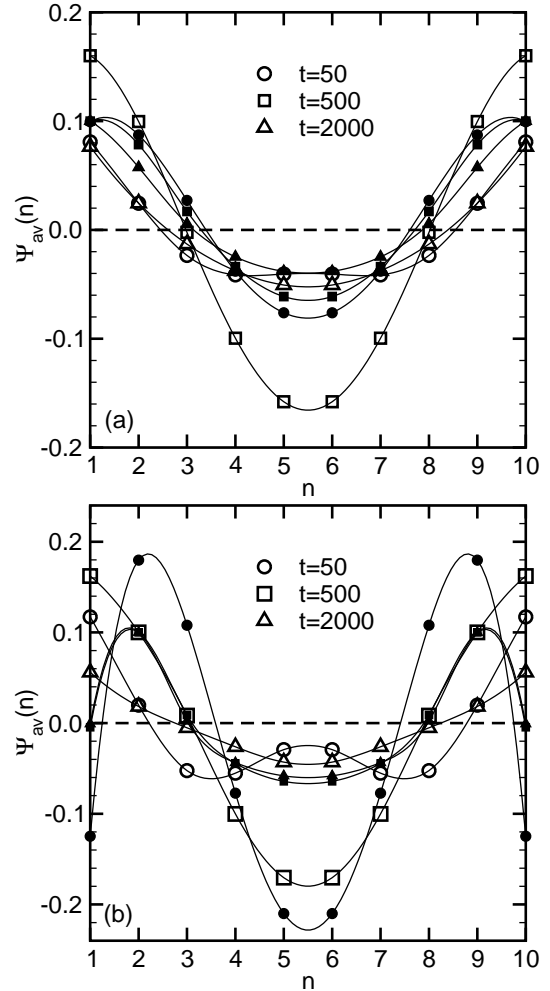

\includegraphics*[width=0.38\textwidth]{fig13a.eps}
\includegraphics*[width=0.38\textwidth]{fig13b.eps}
\caption{\label{fig13} Comparison between the unscaled GL model
[cf. Eq.(\ref{eq28},\ref{eq29})] and lattice model at temperatures
(a) $5J/k_{\rm B}$ and (b) $4J/k_{\rm B}$. Note that the open symbols are for
lattice model and the filled symbols are used for the unscaled 
GL model.}
\end{figure}

From Eq.~(\ref{eq28}) we see, however, that a pathological behavior occurs
if the coefficient $(T_{\rm cb}/T-1-J/T)$ vanishes, which is the case for
$k_{\rm B}T/J=5$: the time evolution of $\phi(\vec{\rho},z=0,t)$ then
is strictly decoupled from the order parameter in the interior, and it
stops if $\phi(\vec{\rho},z=0,t)$ reaches the value $H_{\mathcal{S}1}/J$
(for $J_s/J=1)$. This is what is seen in Fig.~\ref{fig13}(a) where we
have solved the unscaled version of the GL model. In this case $\Psi_{\rm
av}(z=0,t)$ has stopped its time evolution already during the very early
stages. For $k_{\rm B}T<5J$ [Fig.~\ref{fig13}(b)], the coefficient of the
last term on the right side in Eq.~(\ref{eq28}) has changed its sign (in
comparison to the region close to $T_{\rm cb}$), and this leads to the
result that $\phi (\vec{\rho},z=0,t)$ converges to zero, which also is
unreasonable. Thus, taking the coefficient $(T_{\rm cb}/T-1-J/T)$ rather
than simply $(-J/T)$ [the latter leads to the scaled form Eq.~(\ref{eq7})
with the coefficient $\gamma$ as quoted in Eq.~(\ref{eq4})] does not
yield any improvement, but rather is physically inconsistent. However,
working with the scaled form of the GL equations, and their boundary
conditions, Eqs.~(\ref{eq4},~\ref{eq5},~\ref{eq6},~\ref{eq7},~\ref{eq8})
does not yield results in agreement with the lattice model at temperatures
$k_{\rm B}T/J\leq 5$ either.

\section{Conclusion}

In this paper we have presented a Molecular Field Theory for the Kawasaki
spin-exchange Ising model in a thin film geometry and have shown that the
numerical solution of the resulting set of coupled ordinary differential
equations describing the time-dependence of the local magnetization at
the lattice sites is a convenient and efficient method to study spinodal
decomposition of such systems, taking the boundary conditions at the
surfaces of the film properly into account. Obviously, in comparison to
a Monte Carlo simulation of this model one has lost thermal statistical
fluctuations, except for those built into the theory via the choice of
noise in the initial configuration of the system; but there is consensus
\cite{1,2,3,4,5} that for a description of the late stage coarsening
behavior such thermal fluctuations may safely be neglected. Thus, the
present method is favorable in comparison with Monte Carlo, since the
code runs much faster.

In the vicinity of the critical point, where the correlation length $\xi$
is sufficiently large, and also the film thickness $D$ is sufficiently
large as well, our treatment becomes equivalent to the time-dependent
Ginzburg-Landau theory. However, one needs to go surprisingly close to
the critical point of the bulk to actually demonstrate this limiting
behavior from our lattice model treatment numerically. The GL treatment
over most of the parameter regime provides only a qualitative, rather than
quantitative, description of the system. In principle, for the GL theory
to be accurate, the correlation length should be much larger than the
lattice spacing. However, this happens only very close to the critical
point. Away from the critical points no well-defined connection to the
parameters of a microscopic Hamiltonian can be made for the GL theory,
while the present lattice approach has this connection by construction.

Of course, the problem of surface-directed spinodal decomposition is most
interesting for liquid binary mixtures, for which our lattice model is
inappropriate due to the lack of hydrodynamic interactions; even for solid
mixtures (two different atomic species sharing the sites of a lattice) our
model is an idealization (neither elastic distortions nor lattice defects
were included; actual solid binary mixtures decompose via the vacancy
mechanism of diffusion \cite{1,2}; etc.). While GL models including
hydrodynamic interactions have been formulated \cite{1,2,3,4,5}, the
restriction that any GL model is valid only in the immediate neighborhood
of the critical point applies there as well. While outside the critical
region unmixing of fluids can be simulated by Molecular Dynamics methods
(see e.g.~\cite{19}), such simulations are extremely time-consuming,
and an analog of the present dynamic mean field theory for inhomogeneous
fluids would be very desirable. Developing such an approach clearly is
a challenge for the future.

\underline{Acknowledgments}: This work was supported in part by the
Deutsche Forschungsgemeinschaft (DFG), grant number SFB-TR6/A5. K.B. is
very much indebted to Prof. S.  Puri and the late Prof. H.L. Frisch
for stimulating his interest in these problems and for many
discussions. S.K.D. is grateful to K.B. for supporting his stay in Mainz,
where this work was initiated and acknowledges useful discussions with
Prof. S. Puri.


\end{document}